\documentclass[aps,preprintnumbers]{revtex4}

\usepackage{psfrag,graphicx}
\usepackage{dcolumn}
\usepackage{amsmath,amssymb}
\usepackage{bm}
\usepackage{amsfonts,amssymb,amsmath}        

\newcommand{\be}{\begin{equation}}
\newcommand{\ee}{\end{equation}}
\newcommand{\bq}{\begin{eqnarray}}
\newcommand{\eq}{\end{eqnarray}}
\newcommand{\Sp}{\,\,\,\,\,\,}
\newcommand{\no}{\nonumber\\}

\newcommand{\ket}[1]{\left | \, #1 \right\rangle}

\begin{document}

\title{The Wavefunction of an Anyon}
\date{\today}

\author{Jiannis K. \surname{Pachos}}
\affiliation{Department of Applied Mathematics and Theoretical Physics,
University of Cambridge, Wilberforce Road, Cambridge CB3 0WA, UK}

\begin{abstract}

We consider a two-dimensional spin system in a honeycomb lattice
configuration that exhibits anyonic and fermionic excitations [Kitaev,
cond-mat/0506438]. The exact spectrum that corresponds to the
translationally invariant case of a vortex-lattice is derived from which the
energy of a single pair of vortices can be estimated. The anyonic properties
of the vortices are demonstrated and their generation and transportation
manipulations are explicitly given. A simple interference experiment with
six spins is proposed that can reveal the anyonic statistics of this model.

\end{abstract}

\maketitle

\section{Introduction}

Lately, a remarkable effort has been made towards the understanding of
topological systems (see e.g. \cite{Kitaev,Preskill,Freedman} and references
therein). This ranges from the realization of systems that exhibit
topological behavior~\cite{Tsui,Picciotto,Xia} to the systematic
identification and characterization of their topological properties
~\cite{Kitaev_97,Moore,Read,Slingerland,Cooper,Doucot,Freedman}. These
efforts have been greatly boosted by an interest in performing error-free
quantum computation~\cite{Kitaev_97,Freedman1}. The idea is to take
advantage of the non-trivial statistical properties of anyonic particles,
that deviate from the bosonic or fermionic behavior, in order to encode and
process information. In these systems, quantum computation is protected from
any local errors that do not destroy the nature of the anyons.

This article considers in detail a specific two dimensional lattice model
presented by Kitaev~\cite{Kitaev}. It comprises of spin-1/2 particles in a
honeycomb configuration that are subject to nearest neighbor interactions.
The corresponding quadratic Hamiltonian has been solved analytically in the
vortex-free sector and a perturbation treatment was given for its vortex
sector~\cite{Kitaev}. Here we present a non-perturbative study of the
vortex-lattice sector. The latter consists of vortices placed at each
hexagonal plaquette of the lattice. From these analytic results one can
estimate the spectrum of two individual vortices in the limit that the
interactions between vortices are negligible. Our findings are supported by
numerical diagonalization of a small lattice system with 16 spins. A
detailed presentation is given of the manipulation procedures required to
generate or transport the anyons. The considered system can be employed to
perform robust quantum computation with abelian anyons~\cite{Pachos}.

This article is organized as follows. In Section \ref{The_model} we present
the model and the symmetry properties that enable its analytic treatment. In
Section \ref{Spectrum} the separation of the spectrum into different vortex
sectors is presented and the properties of anyonic vortices are elaborated.
The particular cases of the vortex-free and vortex-lattice sectors are
considered and their spectrum is explicitly derived. In the conclusions, the
properties of the spectrum are analyzed and an exact numerical treatment is
presented that supports our findings. Finally, a minimal lattice cell with
six spins is presented that supports anyonic excitations. A simple
interference setup is proposed that can reveal the anyonic statistics
experimentally.

\section{Presentation of the model}
\label{The_model}

Consider a two dimensional system with spin-1/2 particles located at the
vertices of a honeycomb lattice \cite{Kitaev}. The spins are assumed to
interact with each other via the following Hamiltonian
\be
H = -J_x \sum_{\text{x-links}} \sigma_j^x \sigma_k^x
-J_y\sum_{\text{y-links}} \sigma_j^y \sigma_k^y -J_z\sum_{\text{z-links}}
\sigma_j^z \sigma_k^z,
\label{Ham}
\ee
where ``x-links", ``y-links" and ``z-links" are depicted in Figure
\ref{excitations}. Here we shall adopt periodic boundary conditions
although open boundary conditions can be similarly treated. Consider an
individual hexagonal plaquette and the corresponding operator
\be
\hat{w}_p = \sigma_1^x \sigma_2^y \sigma_3^z \sigma_4^x
\sigma_5^y \sigma_6^z,
\label{wp}
\ee
where $\sigma^\alpha_i$ is a Pauli operator at vertex $i$ of the hexagon.
This operator commutes with the Hamiltonian, $H$, as well as with the
corresponding operators, $\hat{w}_{p^\prime}$, of all the other plaquettes.
Thus, its eigenvalues, $w_p=\pm1$, are conserved quantities. Hence we can
split the Hilbert space in sectors, ${\cal L}_w$, characterized by a fixed
configuration of $w_p$'s and solve for the eigenvalues of the Hamiltonian in
each sector. A plaquette with $w_p=-1$ corresponds to a vortex with anyonic
statistics as we shall see in the following. Here the exact diagonalization
of the Hamiltonian will be presented for the limiting cases, where $w_p=1$
or $w_p=-1$, for all plaquettes, $p$. They correspond to the vortex-free and
a regular vortex-lattice configuration, the latter having a vortex at each
hexagonal plaquette.

To diagonalize Hamiltonian (\ref{Ham}) we rewrite the spin operators in
terms of Majorana fermions. The latter are defined as the ``real" and
``imaginary" parts of usual fermionic operators, $a_k$ and $a_k^\dagger$, in
the following way
\be
c_{2k-1} \equiv a_k+a_k^\dagger, \Sp c_{2k} \equiv {a_k-a_k^\dagger \over
i}.
\ee
The Hermitian operators, $c_i$, satisfy the following relations,
$c_i^\dagger = c_i$ and $\{c_i, c_j\} = 2 \delta_{ij}$. Here we introduce
the fermionic picture by representing each spin with two fermionic modes,
$a_1$ and $a_2$ in such a way that the no-fermion state represents a spin up
and the two-fermion state represents a spin down. There are four Majorana
operators that correspond to each spin given by
\be
c_0 = a_1 +a_1^\dagger = c ,\Sp c_1 = {a_1 - a_1^\dagger \over i} = b^x,
\Sp c_2 = a_2 +a_2 ^\dagger = b^y, \Sp c_3 ={a_2 -a_2^\dagger \over i} =
b^z.
\ee
It is possible to project the full space of states of the two fermions to
the physical space ${\cal L}$ of the spin states by employing the projection
operator $D=b^xb^y b^z c$, where
\be
\ket{\Psi} \in {\cal{L}} \Leftrightarrow D\ket{\Psi} = \ket{\Psi}.
\label{constraint}
\ee
Within the subspace ${\cal L}$ the following identifications hold
\be
\sigma^\alpha = i b^\alpha c, \Sp (\alpha = x, y, z),
\ee
where $\sigma^\alpha$ is a specific representation of the Pauli algebra.

To employ this representation for Hamiltonian (\ref{Ham}) we introduce four
Majorana fermions $b^x_j$, $b^y_j$, $b^z_j$ and $c_j$ for each site $j$. As
$[\sigma^\alpha_j,D_k] = 0$ the diagonalization of Hamiltonian ({\ref{Ham})
is re-expressed as diagonalization of $H(\!\{\sigma^\alpha\})=
H(\!\{ib^\alpha_j c_j\})$ accompanied by the constraint $D_j \ket{\Psi} =
\ket{\Psi}$. With the above transformation we have that $\sigma^\alpha_j
\sigma^\alpha_k = -i \hat{u}_{jk} c_j c_k$ with $\hat{u}_{jk} \equiv
ib^\alpha_j b^\alpha_k$, finally obtaining
\be
H = {i \over 4} \sum_{j,k} \hat{A}_{jk} c_j c_k, \Sp \text{where} \Sp
\hat{A}_{jk}= \left\{\begin{array}{c} 2J_\alpha \hat{u}_{jk},
\\ 0, \end{array}
\begin{array}{l} \text{if $j$ and $k$ are connected by a link,}\\
\text{otherwise.} \end{array}
\right.
\label{Ham1}
\ee
Observe that $[H,\hat{u}_{jk}]=0$, $\hat{u}^\dagger_{jk} = \hat{u}_{jk}$ and
$\hat{u}_{jk}^2 = 1$, hence, one can restrict to an eigenspace of states
given by a certain configuration of eigenvalues ($u_{jk}=\pm1$) of the
$\hat{u}_{jk}$ operator. This reduces the diagonalization of Hamiltonian
(\ref{Ham1}) to the diagonalization of the corresponding Hamiltonian, where
$\hat{A}_{jk}$ is substituted by a certain configuration $A_{jk}$ obtained
by replacing $\hat{u}_{jk}$ with its chosen eigenvalue. Thus, instead of
considering the decomposition of the Hilbert space with respect to
eigenvalues of $\hat{w}_p$ one can consider the decomposition, ${\cal L}_u$,
with respect to the eigenstates of $\hat u_{ij}$. Restricted in the physical
subspace ${\cal L}$ the following relation holds
\be
\hat{w}_p = \prod_{(j,k)\in\, boundary(p)}\hat{u}_{jk}
\label{bdry}
\ee
where the $(j,k)$ links are ordered in a clockwise fashion around the
plaquette boundary. However, the operators $\hat{u}_{jk}$ and $D_j$ do not
commute so the resulting eigenstates, $\ket{\Psi_u}$, of the Hamiltonian do
not satisfy the constraint (\ref{constraint}). This is remedied by the
symmetrization
\be
\ket{\Psi_w} = \prod_j \left( {1 +D_j \over 2} \right) \ket{\Psi_u}.
\label{symm}
\ee
Thus, the eigenstates of Hamiltonian (\ref{Ham1}) have to be symmetrized
with the above procedure in order to obtain the physical states that
correspond to the vortex configurations of Hamiltonian (\ref{Ham}).
Obviously this procedure does not affect the eigenvalues of the Hamiltonian.

\section{Spectrum of Hamiltonian}
\label{Spectrum}

\subsection{Sectors of Hilbert space}

As we have seen, it is possible to reduce the full Hilbert space of
Hamiltonian (\ref{Ham}) to independent subspaces, ${\cal L}_w$, that
correspond to different eigenvalue configurations of $\hat w_{p}$. The
subspace with $w_p=+1$ for all plaquettes $p$ corresponds to the vortex-free
configuration. Diagonalizing Hamiltonian (\ref{Ham1}) in this subspace will
just give a fermionic spectrum with no anyons (vortices) present. Changing
the sign of one $u_{jk}$ results in two adjacent plaquettes having $w_p=-1$.
This configuration corresponds to two vortices placed at these plaquettes.
In the absence of  external fields these anyonic excitations are static in
the sense that they remain there indefinitely. One can perform this change
of sign in $u_{jk}$ by simple Pauli rotations of the original spins. Indeed,
a $\sigma^z$ rotation on a certain spin will cause the generation of two
vortices in the adjacent plaquettes, as shown in Figure~\ref{excitations}.
Hamiltonian (\ref{Ham1}) can now be diagonalized in this subspace of states
which results into a fermionic spectrum. The ground state of each sector
corresponds to the pure anyonic configuration without fermions.

\begin{center}
\begin{figure}[ht]
\resizebox{!}{3 cm}
{\includegraphics{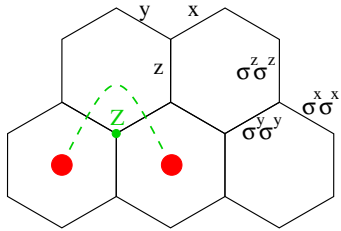}}
\caption{\label{excitations} The honeycomb lattice with the three different
types of links. At each type corresponds a different spin-spin interaction.
Two vortex configuration is induced by a $\sigma^z$ rotation at a certain
spin. This rotation does not commute with the interaction terms on the $x$-
and $y$-links of Hamiltonian (\ref{Ham}), a property represented by the
dotted connecting string.}
\end{figure}
\end{center}

To study the vortex excitations as well as their transport properties
systematically we shall adopt the operators that correspond to the toric
code limit of the model \cite{Kitaev,Pachos} obtained, e.g. when $J_z\gg
J_x,J_y$. In this limit there are three vortex excitations above the ground
state, $\ket{gs}$, given by
\be
\ket{Z} \equiv\sigma^z\otimes 1 \ket{gs}\approx 1\otimes \sigma^z \ket{gs},
\Sp \ket{Y} \equiv \sigma^x\otimes \sigma^y\ket{gs}\approx \sigma^y
\otimes \sigma^x \ket{gs}, \Sp \ket{X} \equiv i\sigma^x \otimes \sigma^x
\ket{gs}.
\ee
The two operators of the tensor product are acting on the two spins of a
certain $z$-link. The $\ket{Z}$ and $\ket{Y}$ excitations behave as bosons
with themselves while the $\ket{X}$ excitation is fermionic. The latter can
be produced by fusing the $\ket{Z}$ and $\ket{Y}$ particles and its energy
is the sum of the energies of the two constituents. Appropriate applications
of the Pauli operators result in the transport of the anyonic subspaces
around the lattice. All of the excitations behave as anyons with each other
with a statistical angle $\theta =\pi/2$ that results from the
anticommutation relations of the Pauli operators. Similarly, the fusion
properties of the vortices follow from the properties of the Pauli
operators, $X\times X=Y\times Y= Z\times Z=1$, $X\times Y= Z$ and
permutations, where $1$ is the no-vortex configuration. For general values
of the couplings $J_\alpha$ these excitations may be accompanied by
non-vortex fermions that reside at each sector. This, of course, does not
affect the anyonic properties of the vortices. Hence, it is possible to move
between different subspaces, ${\cal L}_w$, by applying Pauli rotations on
the original spins.

The generation and transport properties of the different sectors do not
necessarily correspond to the properties of the anyonic excitations without
the fermions. In general, a simple $\sigma^z$ rotation moves the vortex-free
configuration to the one with two vortices, but fermions may also be
generated. To guarantee that no unwanted fermions are created one may
consider realizing locally a regime for which one coupling is much larger
than the other two, e.g. $J_z\gg J_x,J_y$. Then the fermionic energy gap is
much larger than the anyonic one~\cite{Pachos}, while we can still find
Pauli rotations that correspond to the exchange of energy of the order of
the anyonic gap. Indeed, a $\sigma^z$ rotation of spin $i$ of the initial
Hamiltonian (\ref{Ham}) results in a two vortex configuration
\be
H_{\text{2-v}} = \sigma^z_i H\sigma^z_i = H +2J_x \sigma^x_i \sigma^x_j +2J_y
\sigma^y_i\sigma^y_k,
\label{exx}
\ee
where $(i,j)$ denotes an $x$-link and $(i,k)$ denotes a $y$-link, and, hence
it causes an exchange of energy that scales with $J_x$ and $J_y$. As we
shall see in the following, the fermionic excitations are of the order of
$J_z$. Hence, they are decoupled from the rest of the system and are
negligible in manipulations at this energy scale. It is worth noting that
diagonalizing Hamiltonian (\ref{exx}) does not give a $\sigma^z$ rotated
$\ket{Z}$ state (which would correspond to the lowest energy vortex-free
configuration) as that state would correspond to a different sector.

\subsection{The vortex-free sector}

Due to the symmetry of the Hamiltonian we can restrict ourselves to the case
of positive couplings $J_\alpha$ \cite{Kitaev}. The simplest configuration
of the eigenvalues of $\hat u_{jk}$ is given by $u_{jk}=1$ for all links
$(j,k)$ of every hexagonal plaquette taken in a clockwise fashion. This
configuration is also known as the vortex-free configuration as it gives
$w_p=1$ for all plaquettes $p$, and can easily be solved by a Fourier
transformation.

\begin{center}
\begin{figure}[ht]
\resizebox{!}{3.5 cm}
{\includegraphics{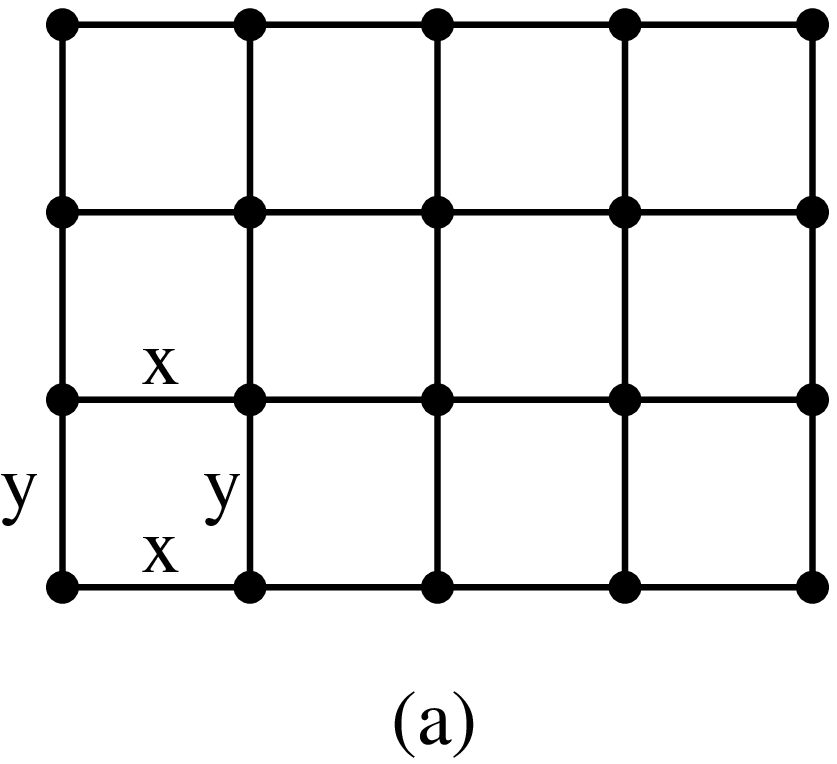} \Sp \Sp \Sp \Sp \Sp \Sp \Sp \Sp
\includegraphics{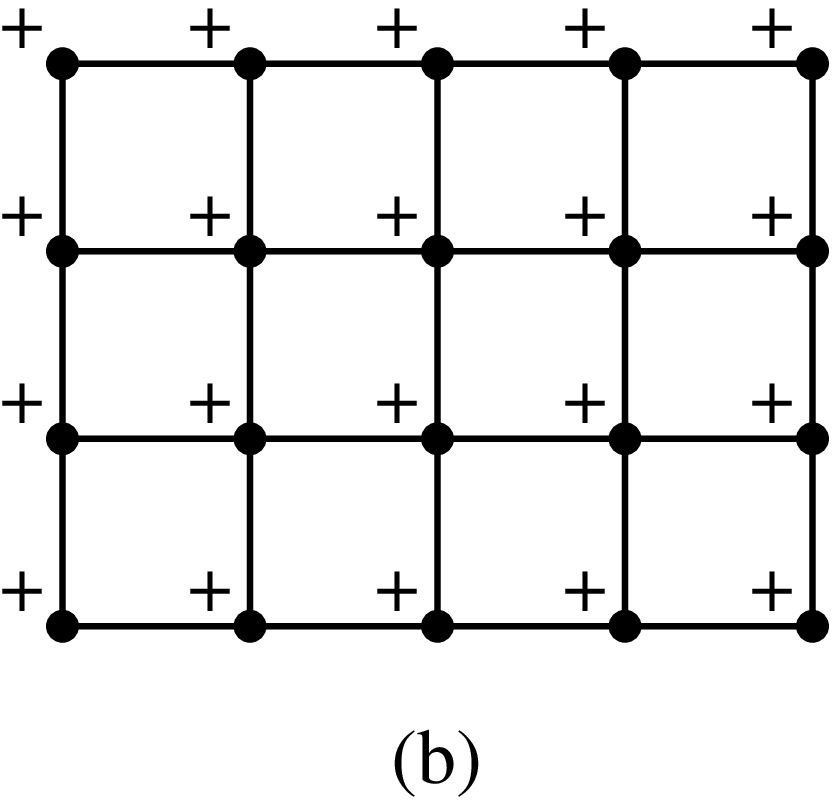}\Sp \Sp \Sp \Sp \Sp \Sp \Sp \Sp
\includegraphics{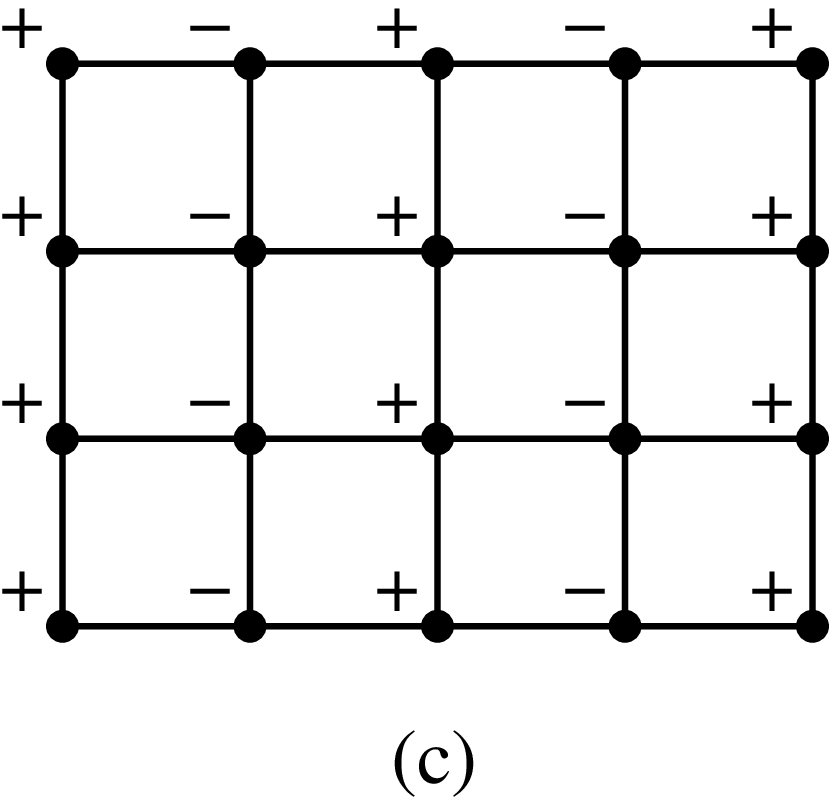}}
\caption{\label{square} (a) The effective square lattice of the original
honeycomb model. Each cell (filled circle) incorporates two vertices of the
original lattice and a $z$-link, while the depicted $x$ and $y$-links have
unit length. (b) The vortex-free case where $u_{jk}=+1$ for $(j,k)$ taken
clockwise around each plaquette. (c) The vortex-lattice where $u_{jk}=+1$
for all $(j,k)$ oriented clockwise except for the ones that correspond to
every other $z$-link in the $x$ direction.}
\end{figure}
\end{center}

It is convenient to rearrange the lattice in the following way. Let a unit
cell include the $(j,k)$ site that corresponds to the $z$-link. Each cell
can be represented by an index $s$ while the position of the vertex within
the cell (i.e. distinguishing between $j$ and $k$) is represented by an
index $\lambda$ (see Figure \ref{square}(a)). Thus, the Hamiltonian takes
the form
\be
H = {i \over 4} \sum_{s\lambda,t \mu} A_{s \lambda,t\mu} c_{s\lambda}
c_{t\mu} .
\label{Ham_square}
\ee
As Hamiltonian (\ref{Ham_square}) is translationally invariant we employ
Fourier transforms to diagonalize it. Let us define
\be
c_\lambda(p) = {1 \over \sqrt{2 N}} \sum_s e^{-i p\cdot s} c_{s \lambda},
\Sp
A_{\lambda \mu}(p) = \sum_t e^{i p \cdot t} A_{0\lambda, t \mu},
\ee
where $p_{x,y}\in[-\pi,\pi]$ and the two dimensional vector, $t=(t_x,t_y)$,
indicates the position of the vertices on the effective square lattice. The
Hamiltonian becomes
\be
H = {i \over 2}\sum_{ \lambda \mu}\int d^2\!p\, A_{\lambda
\mu} c_{\lambda}^\dagger(p) c_{\mu}(p)
\ee
where $c^\dagger_{ \lambda}(p) = c_{ \lambda}(-p)$, $c_{ \lambda}(p)
c_{\mu}^\dagger(q) +c_{\mu}^\dagger(q) c_{ \lambda}(p) = \delta_{pq}
\delta_{\lambda\mu}$ and
\be
i A(p) = \left(\begin{array}{cc} 0 & i f(p) \\
-if^* (p) & 0 \end{array}\right) , \Sp f(p) = 2(J_x e^{ip_x} +J_y e^{ip_y} +
J_z).
\ee
For $\varphi \equiv \arg(f(p))$ one can introduce the fermionic operators
\be
b(p) = \big(c_1(p) + i e^{i\varphi} c_2(p)\big)/\sqrt{2},
\Sp b^\dagger(p) = \big(c_1(-p) - i e^{-i\varphi} c_2(-p)\big)/\sqrt{2}.
\ee
In terms of $b$ the Hamiltonian takes the diagonal form
\be
H = \int d^2\! p\, |f(p)| \left[b^\dagger(p) b(p) -{1 \over 2} \right].
\ee
The ground state is given by
\be
\ket{gs} = \prod_{p} b(p) \ket{0} =
\prod_p \big(c_1^\dagger(p) + i e^{-i \varphi} c_2^\dagger (p)\big) \ket{0},
\ee
where $\ket{0}$ is the vacuum state of the Majorana operators,
$c_\lambda(p)$. The overall ground state energy is $E_{gs} = -{1 \over
2}\int d^2\!p\, |f(p)|$. The first excited state and its energy gap above
the ground state are given by
\be
\ket{e_{p_0}} = b^\dagger(p_0)\ket{0}, \Sp \Delta E_{p_0} =
\min_{p_0} |f(p_0)|.
\ee
It is clear that if $|J_x|$, $|J_y|$, $|J_z|$ satisfy
\be
|J_x|\leq|J_y| + |J_z|, \Sp |J_z|\leq|J_x| + |J_y|, \Sp |J_y|\leq|J_z| +
|J_x|,
\ee
then there exists a $p_0$ such that $\Delta E_{p_0} = 0$, i.e. the gap
vanishes (see Figure~\ref{triangle}(a)). The excitation spectrum of the
operators $b(p_0)$ corresponds to the fermionic spectrum of this sector
($w_p=1$ for all plaquettes, $p$). Assuming that the coupling configuration
is not at the borders of the gapless region one can easily verify that in
the gapless region there are two distinctive momenta for which the energy
gap becomes zero. At these two Fermi points the energy has conical
singularities.

\begin{center}
\begin{figure}[ht]
\resizebox{!}{4 cm}
{\includegraphics{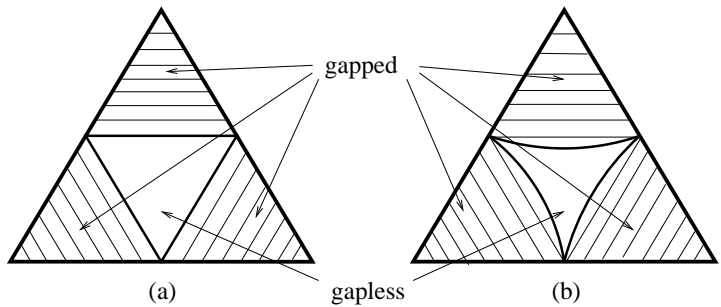}}
\caption{\label{triangle} (a) The gapped and gapless phases for the
vortex-free case. (b) The gapped and gapless phases for the vortex-lattice
case. For both plots $J_x+J_y+J_z=1$ where we assume $J_x,J_y,J_z \ge 0$.
For the vortex-free case (a) the inner boundaries are given by the
conditions $J_x+J_y= J_z$, $J_y+J_z= J_x$ and $J_z+J_x = J_y$, while for the
vortex-lattice (b) the conditions are $J_x^2+J_y^2= J_z^2$, $J_y^2+J_z^2=
J_x^2$ and $J_z^2+J_x^2 = J_y^2$.}
\end{figure}
\end{center}

\subsection{The vortex-lattice sector}

The anyonic spectrum can be obtained by studying the configuration with $w_p
=-1$ for all $p$. To implement it we consider the particular case where
$u_{jk}$ has alternating signs at the $z$-links along the $x$ direction
while it is homogeneous in the $y$ direction (see Figure \ref{square}(c)).
This is sufficient to create $w_p = -1$ for all plaquettes $p$ giving rise
to a vortex-lattice where a vortex is placed at each hexagonal plaquette.

As the sign of the $z$-links coupling is now alternating compared to the
previous vortex-free case, the Hamiltonian in the momentum representation
takes the form
\be
H = {i \over 2} \sum_{ \lambda \mu}\int d^2\!p\, d^2\!p' A_{\lambda \mu}
(p,p') c_\lambda(p)c_\mu(p'),
\ee
where $A_{12} (p,p') = \delta(p+p') M(p) +\delta(p+p'+\pi) N(p)$, $\Sp M(p)
\equiv 2J_x e^{ip_x} + 2J_y e^{i p_y}$, $N(p) \equiv 2 J_z$ and
$A_{21}=-A_{12}^*$. Explicitly we have
\be
H = {i \over 2} \int\int_{-\pi}^{\pi} dp_xdp_y \left\{ M(p) c_1^\dagger(p)
c_2(p)+ M^*(p) c_1(p)c_2^\dagger(p) + N(p)\left[c_1^\dagger(p)
c_2(\tilde{p}) +c_1(p) c_2^\dagger(\tilde{p})\right]\right\},
\label{Ham5}
\ee
where $p=(p_x,p_y)$ and $\tilde{p} = (p_x \pm\pi,p_y)$. As the last term of
(\ref{Ham5}) is repeated twice in the interval $p_x\in[-\pi,\pi]$ the
Hamiltonian can be rewritten as
\be
H = {i \over 2} \int_{-\pi/2}^{\pi/2} dp_x \int_{-\pi}^{\pi}dp_y
\left\{ M(p) c_1^\dagger(p) c_2(p)+ M(\tilde{p})
c_1^\dagger(\tilde{p})c_2(\tilde{p}) + N(p)\left[c_1^\dagger(p)
c_2(\tilde{p}) +c_1^\dagger(\tilde{p}) c_2(p)\right]\right\} +\text{h.c.}.
\label{Ham6}
\ee
To diagonalize the Hamiltonian \cite{Verkholyak} we would like to transform
(\ref{Ham6}) into the following form
\be
H = {i \over 2} \int_{-\pi/2}^{\pi/2} dp_x \int_{-\pi}^{\pi}dp_y \left[A(p)
\eta_1^\dagger(p) \eta_2(p) + B(p) \eta_1^\dagger (\tilde{p})
\eta_2(\tilde{p})\right] +\text{h.c.}.
\label{Ham2}
\ee
The corresponding canonical transformation has to satisfy
\be
[H,\eta_1^\dagger(p)] = -{i \over 2} A^*(p) \eta_2^\dagger(p),
\Sp [H, \eta_1^\dagger(\tilde{p})]= -{i \over 2} B^*(p)
\eta_2^\dagger(\tilde{p}),
\Sp [H,\eta_2^\dagger(p)] = {i \over 2} A(p) \eta_1^\dagger(p),
\Sp [H,\eta_2^\dagger(\tilde{p})] = {i \over 2} B(p)
\eta_1^\dagger(\tilde{p}).
\nonumber
\ee
One can verify that the appropriate transformation is given by
\bq
&& c_\lambda(p) = \cos \theta \eta_\lambda(p) - \sin\theta e^{-(-1)^\lambda
i\phi} \eta_\lambda(\tilde{p}),
\no \no
&& c_\lambda(\tilde{p}) = \sin\theta e^{(-1)^\lambda i\phi} \eta_\lambda(p)
+ \cos \theta \eta_\lambda(\tilde{p}),
\label{trans}
\eq
where $\phi = \arg(M(p) +M^*(\tilde{p}))$ and
\be
\tan 2 \theta = { 2 |M(p) +M^*(\tilde{p})| N \over |M(p)|^2 -
|M(\tilde{p})|^2}.
\ee
The coefficients $A(p)$ and $B(p)$ of Hamiltonian (\ref{Ham2}) are given by
\bq
&& A(p) = {|M(p)|^2+|M(\tilde{p})|^2+2M(p)M(\tilde{p}) +\sqrt{\Delta} \over
M^*(p) + M(\tilde{p})}
\no \no
&& B(p) = {|M(p)|^2+|M(\tilde{p})|^2+2M(p)M(\tilde{p}) -\sqrt{\Delta} \over
M(p) + M^*(\tilde{p})},
\eq
where
\be
\Delta= \big(|M(p)|^2-|M(\tilde{p})|^2\big)^2 +|M(p) + M^*(\tilde{p}))|^2N^2.
\ee
Finally, the Hamiltonian can be diagonalized by employing the following
fermionic operators
\bq
&& b(p) = \big(\eta_1(p) + ie^{i\varphi}\eta_2(p)\big)/\sqrt{2}, \Sp\Sp
b(\tilde p) = \big(\eta_1(\tilde p) +ie^{i\varphi'}\eta_2(\tilde
p)\big)/\sqrt{2},
\no \no &&b^\dagger(p) =
\big(\eta^\dagger_1(p) - ie^{-i\varphi}\eta^\dagger_2(p)\big)/\sqrt{2}, \Sp
b^\dagger(\tilde p) = \big(\eta^\dagger_1(\tilde p)
-ie^{-i\varphi'}\eta^\dagger_2(\tilde p)\big)/\sqrt{2}.
\eq
where $\varphi=\arg(A(p))$ and $\varphi'=\arg(B(p))$, eventually giving
\be
H =  \int_{-\pi/2}^{\pi/2} dp_x \int_{-\pi}^{\pi}dp_y\left[ |A| b^\dagger
(p) b(p) + |B| b^\dagger(\tilde p) b(\tilde p) - {|A| +|B| \over 2}\right].
\ee
The ground state with its energy is given by
\be
\ket{\text{gs}} = \prod_{- \pi/2 <q<\pi/2} b(q) b(\tilde q) \ket{0},
\Sp E^{\text{v-l}}_{\text{gs}} =- \int_{-\pi/2}^{\pi/2} dp_x
\int_{-\pi}^{\pi}dp_y
{|A|+|B| \over 2},
\ee
while some of the first excited states with their energy gap are given by
\bq
&&
\ket{e_{p_0
}} = b^\dagger (p_0) \ket{\text{gs}}, \Sp \Delta E_{p_0} =
\min_{p_0}|A(p_0)|,
\no \no
&&
\ket{e_{\tilde p_0}} = b^\dagger (\tilde p_0) \ket{\text{gs}},
\Sp \Delta E_{\tilde p_0} = \min_{p_0}|B(\tilde p_0)|.
\eq

It is possible to derive the values of the couplings for which the
vortex-lattice configuration becomes gapless. Note first that $|A(p)|$
cannot be zero, but $|B(p)|$ can. Indeed, there exists $p$ such that
$|B(p)|=0$ if and only if all of the conditions below are satisfied
\be
J_x^2 + J_y^2 \ge J_z^2, \Sp J_y^2 + J_z^2 \ge J_x^2, \Sp J_z^2 + J_x^2 \ge
J_y^2,
\label{gapless_phase}
\ee
which is given schematically in Figure \ref{triangle}(b). When
(\ref{gapless_phase}) is satisfied and for coupling configurations away from
the borders of the gapless region one can verify that there are two
distinctive momenta for which the energy gap becomes zero. At these two
Fermi points the energy has conical singularities.

\section{Conclusions}
\label{Conclusions}

As we have seen, vortices with respect to the $u_{jk}$ gauge field have
anyonic statistics and are generated in the system by applying proper spin
rotations. They are connected either with other vortices, or, when the
system is open, with its boundary by string operators. The vortex
excitations are static, but they can be transported by applying external
fields. Each vortex configuration is accompanied by a fermionic spectrum. We
evaluated this spectrum for the limiting cases of the vortex-free and the
vortex-lattice configurations. Their ground states are fermionic vacua (pure
anyons) from which one can obtain the physical wavefunctions by performing
symmetrization (\ref{symm}).

It is of interest to study the behavior of the energy of these
configurations as functions of the couplings $J_\alpha$. When $J_x=J_y=0$
then $E^{\text{v-f}}_{\text{gs}} = E^{\text{v-l}}_{\text{gs}} =-J_z N$ where
$N$ is the total number of plaquettes for both the vortex-free and the
vortex-lattice configurations. It is possible to check that for all other
values of the couplings $J_\alpha$ we have $E^{\text{v-f}}_{\text{gs}}\leq
E^{\text{v-l}}_{\text{gs}} $. Hence, the energy minimum is achieved by the
vortex-free configuration, a fact that also follows from a theorem by Lieb
\cite{Lieb} and has been verified numerically by Kitaev
\cite{Kitaev}. It is easily observed that the two energies become equal
only when one of the couplings $J_\alpha$ becomes zero.

One can compare the fermionic excitation energy of the vortex-free
configuration with the energy gap of a pair of vortices. To derive the
energy of this pair from the energy of the vortex-lattice we assume that the
vortices are non-interactive. This assumption is valid in the gapped regime
where any interaction mediated by the fermions is exponentially
suppressed~\cite{comment}. This is also supported by exact numerical
diagonalization as we see in the following. In Figure
\ref{energy_gaps}(a) the energy gap is plotted between the ground state and
the anyonic  or the fermionic excitations. One can easily see that there is
a transition in the character of the first excitation from anyonic to
fermionic.
\begin{center}
\begin{figure}[ht]
\resizebox{!}{6 cm}
{(a)\includegraphics{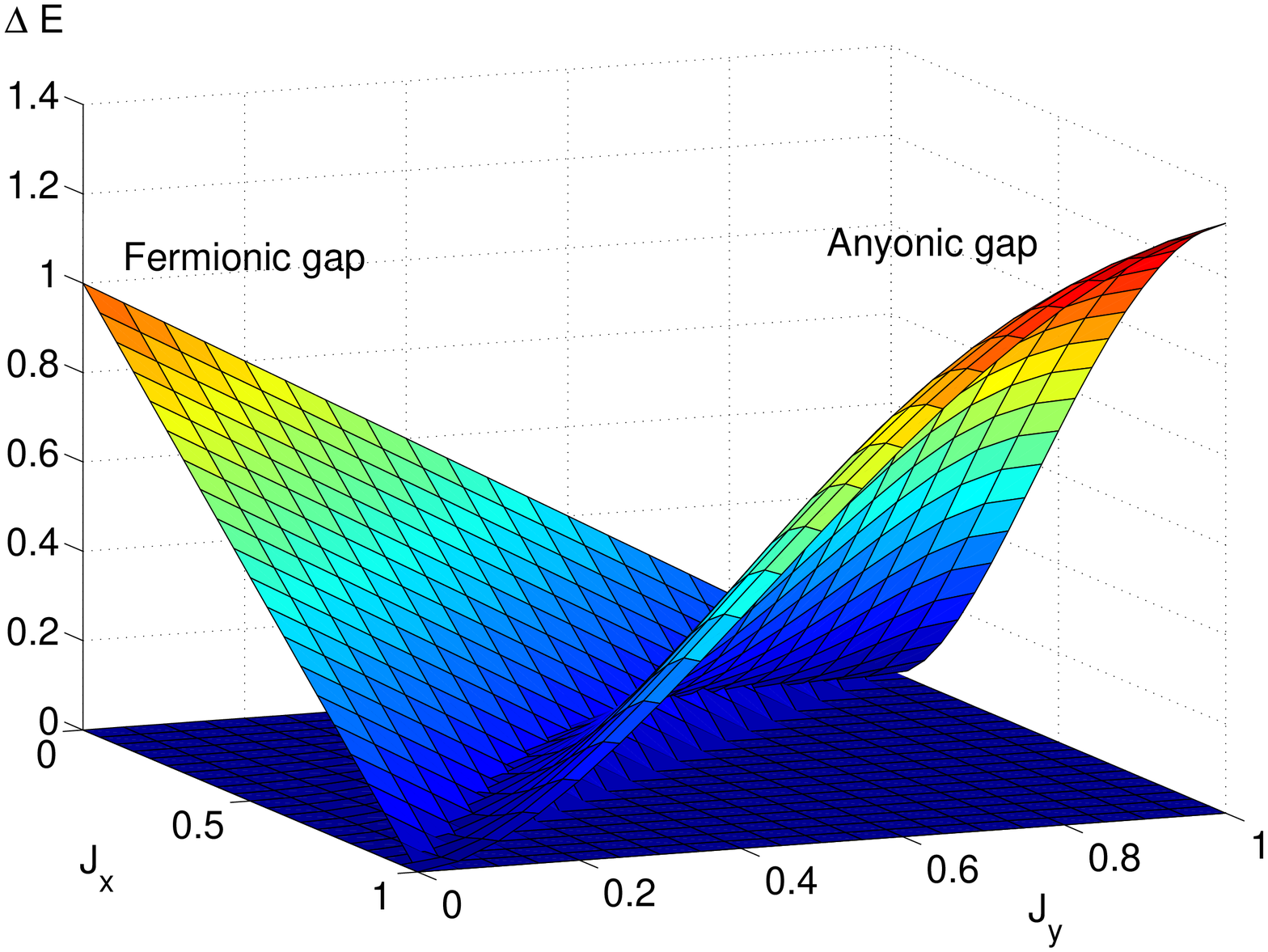}}
\,\,\,\,\,\,\,\,\,\,\,
\resizebox{!}{6 cm}
{(b)\includegraphics{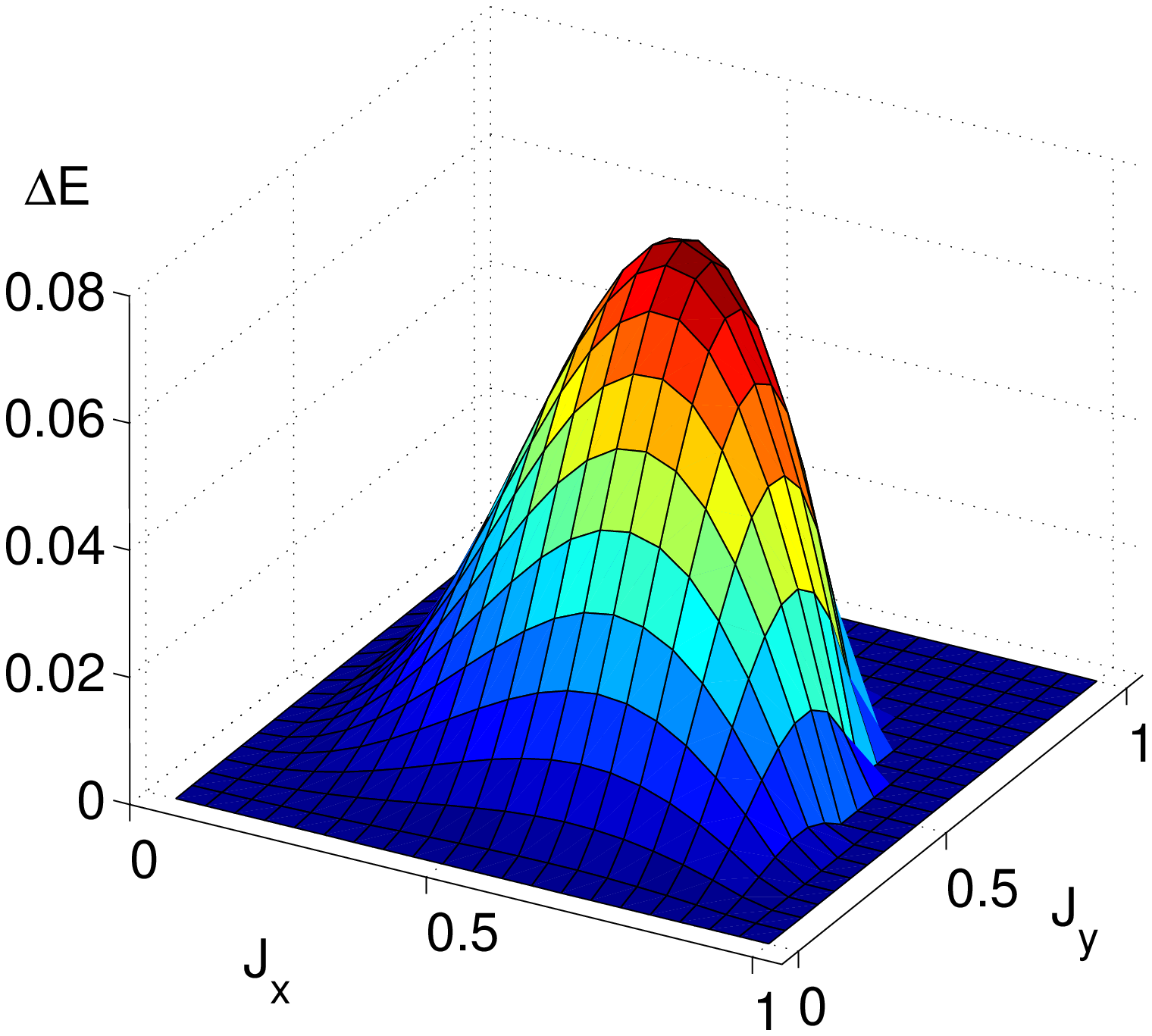}}
\caption{\label{energy_gaps} (a) Analytic results for the energy of
a pair of anyons and of a fermionic excitation. The anyonic and fermionic
energy gaps are plotted against $J_x$ and $J_y$ where we have set $J_z=1$.
(b) Numerical evaluation of the energy gap of the first excited state above
the ground state. This gap is in agreement with the one given as a minimum
of the vortex and fermionic gaps as seen in (a).}
\end{figure}
\end{center}
To verify our findings we simulate the original Hamiltonian with 16 spins in
a periodic honeycomb lattice. In Figure \ref{energy_gaps}(b) we plot the
energy gap between the ground state and the first excited one against $J_x$
and $J_y$, while keeping $J_z=1$. There we see that the excitation becomes
gapless for values of the couplings which agree with the behavior of the
vortex and fermionic excitations. Indeed, the energy gap obtained
numerically is approximately nested  below the analytically obtained energy
gaps with a maximum value that agrees well with the value obtained from
Figure
\ref{energy_gaps}(a).

We conclude by presenting a simple model with six spins that can reveal the
anyonic statistics of the excitations experimentally. It consists of one
hexagonal plaquette with only the spin interactions that lay on the hexagon
(see Figure~\ref{plaquette}). An anyonic excitation in the plaquette can be
produced by a $\sigma^z_6$ rotation. Its paired anyon is outside our system
thus it does not increase the total energy of the system. In order to obtain
the anyonic properties one would like to circulate another anyon around it.
Even if it is not possible to have two anyons present in this simple model,
one can consider anyons outside the system that can circulate the existing
anyon by appropriately rotating the six spins. Indeed, performing the
operation $S =\sigma^z_1 \sigma^y_2\sigma^x_3
\sigma^z_4 \sigma^x_5 \sigma^y_6$ corresponds to a looping trajectory around
the anyonic excitation in the plaquette and, thus, it should generate an
overall $\pi$ phase factor. Indeed, for the ground state, $\ket{gs}$, and
the vortex state, $\ket{v}=\sigma^z_6\ket{gs}$, this evolution is given by
\be
S\ket{v} = S \sigma^z_6 \ket{gs} = - \sigma^z_6 S\ket{gs} = -\sigma^z_6
\ket{gs} = -\ket{v}
\ee
The property $S\ket{gs}=\ket{gs}$ is due to the invariance of the ground
state under closed loop operations. It is worth noting that $S$ is also the
$\hat{w}_p$ operator (\ref{wp}) for the single plaquette of our system.

\begin{center}
\begin{figure}[ht]
\resizebox{!}{2.7 cm}
{\includegraphics{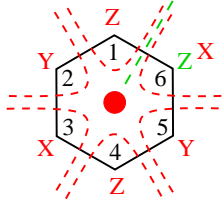}}
\caption{\label{plaquette} A hexagon that can support anyonic
statistics. A vortex is generated, e.g. by a $\sigma^z$ rotation of spin
$6$. Pauli rotations around the plaquette result in the generation of the
looping trajectory. This procedure results in a $\pi$ phase revealing the
anyonic statistics of the vortex.}
\end{figure}
\end{center}

This procedure can be used in an interferometric setup to observe the
anyonic statistics experimentally. Indeed, if one performs the rotation
$R=(1-i\sigma^z_6)/\sqrt{2}$, then the superposed state $(\ket{gs}
-i\ket{v})/\sqrt{2}$ is produced. The looping evolution and the inverse
rotation $R^{-1}$ will result into the $\ket{v}$ state if the statistics of
the vortex excitation is anyonic, while a bosonic or fermionic statistics
would give back the $\ket{gs}$ state. This interference effect produced with
only one hexagonal plaquette can, in principle, be realized with present
technology using NMR~\cite{Glaser} or Josephson Junctions
\cite{Ioffe}. Alternatively, one can consider a lattice of vortices where
every other vortex is transported around its static neighbor. Such an
experiment could be realized in an optical lattice setup with
atoms~\cite{Duan} or polar molecules~\cite{Zoller}. The readout of the final
configuration could be performed by energy addressing using the spectrum
derived in Section~\ref{Spectrum}.

\acknowledgements

The author would like to thank Alexei Kitaev and John Preskill for inspiring
conversations, Roger Colbeck for critical reading of the manuscript and KITP
for its hospitality. This research was supported in part by the National
Science Foundation under Grant No. PHY99-0794 and by the Royal Society.


\begin{thebibliography}{7}
\bibitem{comment} In the fermionic gapless phase we expect
the anyons to be dressed with fermions that propagate interactions between
them and change their energy significantly.
\end{thebibliography}
\end{document}